\documentclass[pra,twocolumn,preprintnumbers,floatfix]{revtex4}
\usepackage{graphicx}
\usepackage{dcolumn}
\usepackage{bm}
\usepackage{latexsym,epsfig}
\usepackage{graphicx}
\usepackage{verbatim}
\usepackage{comment}
\usepackage{amsmath}
\usepackage{amssymb}
\usepackage{stmaryrd}
\usepackage{color}
\usepackage{epstopdf}
\usepackage{grffile}
\usepackage{lipsum}
\DeclareGraphicsExtensions{.eps}

\newcommand{\beq}{\begin{equation}}
\newcommand{\eeq}{\end{equation}}
\newcommand{\bea}{\begin{eqnarray}}
\newcommand{\eea}{\end{eqnarray}}
\newcommand{\ben}{\begin{eqnarray*}}
\newcommand{\een}{\end{eqnarray*}}
\newcommand{\bfig}{\begin{figure}}
\newcommand{\efig}{\end{figure}}

\newcommand{\ra}{\rangle}

\newcommand{\ket}[1]{|{#1}\rangle}

\begin{document}
\title{Emergent photon pair propagation in circuit QED with superconducting processors}
\author{Sayan Lahiri, Suman Mondal, Kanhaiya Pandey and Tapan Mishra}
\affiliation{Department of Physics, Indian Institute of Technology, Guwahati-781039, India}
\date{\today}

\begin{abstract}
We propose a method to achieve photon pair propagation in an array of three-level superconducting circuits. Assuming experimentally 
accessible three-level artificial atoms with strong anharmonicity coupled via microwave transmission lines in both one and two dimensions we analyze the 
circuit Quantum Electrodynamics(QED) of the system. We explicitly 
show that for a suitable choice of the coupling ratio between different levels, 
the single photon propagation is suppressed and the propagation of photon pairs emerges. 
This propagation of photon pairs leads to the pair superfluid of polaritons associated to the system. 
We compute the complete phase diagram of the polariton quantum matter revealing the 
pair superfluid phase which is sandwiched between the vacuum and the Mott insulator  state corresponding 
to the polariton density equal to two in the strong coupling regime.
\end{abstract}

\maketitle

The phenomenon of pairing plays significant roles in different areas of fundamental physics ranging from condensed 
matter to atomic, molecular and nuclear physics. Typically, in such systems the two body attractions lead to the 
formation of bound states of constituent particles. However, in some specific cases the bound pairs can be formed
even in the presence of two particle repulsion e.g. the cooper pairs of electrons~\cite{ashcroft2011solid} or Superconductor.
The two-body interactions whether attractive or repulsive lead to the formation of bound states of constituent particles.  
These pairs under proper conditions, may have significant contributions in 
establishing novel and exotic physical phenomena and contribute to technological applications. 
In recent years the simplest such pair formations(attractive and repulsive) 
have been predicted and experimentally observed in the context of interacting ultracold atomic systems 
in optical lattices~\cite{baranovprl,doublon,repulsivepair}. These observations relies on the sophisticated control 
over the parameters associated with the optical lattice strength and/or 
the technique of Feshbach resonance~\cite{stoof}.
Although, the atomic or molecular systems 
provide promising platforms to simulate several complex quantum many-body phenomena, 
there are certain limitations which can not be avoided due to various reasons.   
In particular, the formation of attractive pairs will require three-body 
hardcore constraint which involves three-body inelastic losses~\cite{baranovprl} resulting in
extremely small life time of the atomic pairs. 
On the other hand the formation of Feshbach molecules are rovibrationally unstable 
and can reduce to the lower levels very easily. At this point it is believed that the interacting photons 
can form stable bound pairs which can provide 
promising platform to explore various fundamental phenomena and further the scope for technological applications. Several successful attempts have been made to 
create bound states of photons under different conditions~\cite{vuletic1,vuletic2,pnas}. The primary thrust and interest in creating photonic bound states rests not only 
to understand the fundamental physics of nature but also on possible practical applications in quantum communications and technologies. 

\begin{figure}[t]
\centering
  \includegraphics[width=1.0\linewidth]{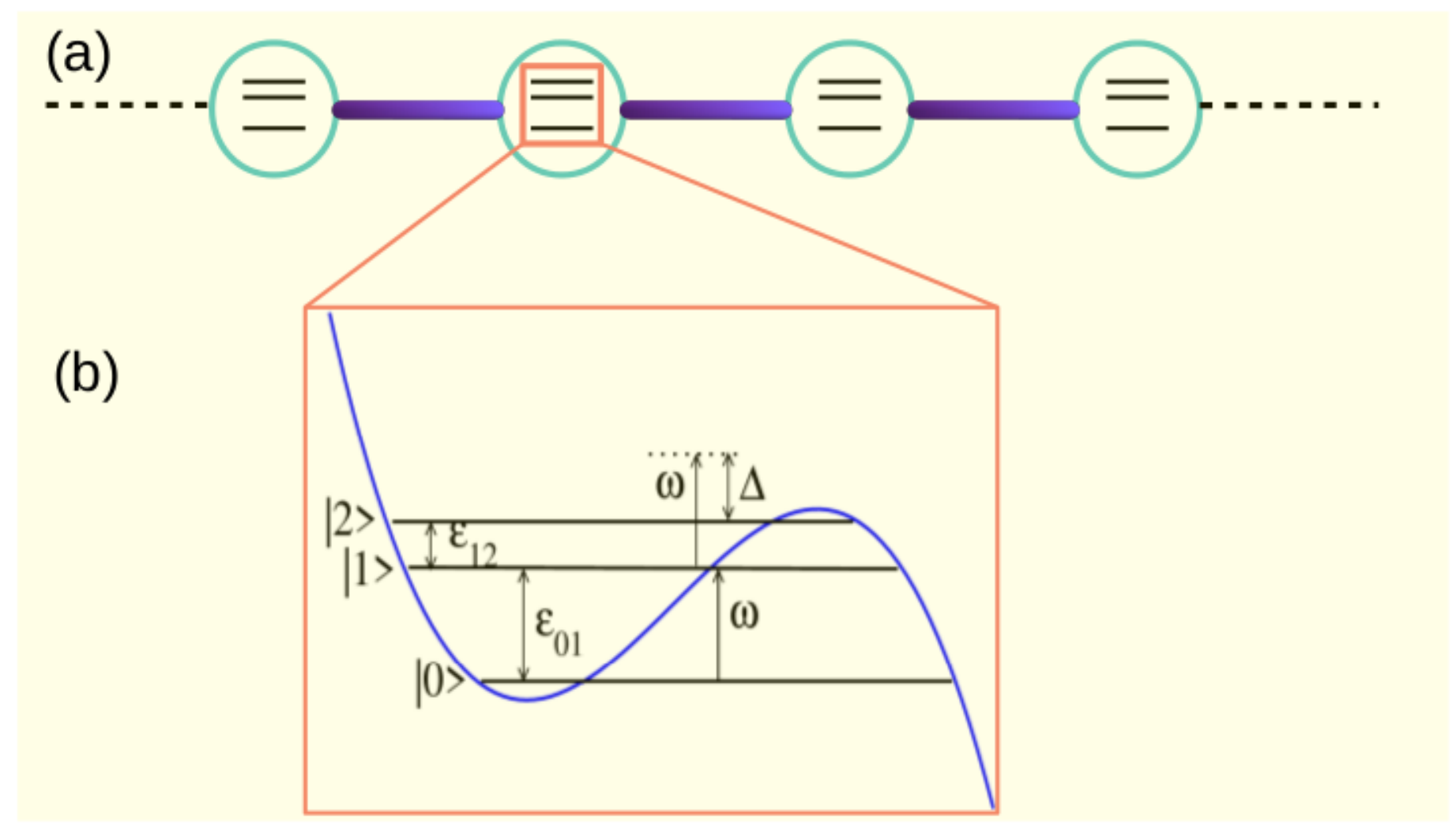}
\caption{(Color online) (a)Coupled cavity array with each cavity containing a SQC which behaves as a three level artificial atom with unequally spaced energy levels. 
(b)Shows the energy levels with only one driving frequency $\omega$. We set $\omega=\varepsilon_{01}$ ($\hbar$=1) and $\Delta=\omega-\varepsilon_{12}$.}
\label{fig:scheme}
\end{figure}

The realization of strong interaction between photons has been a topic of paramount interest in last several decades. 
The interaction which is believed to exist in optical 
non-linear media however, does not possess enough non-linearity to ensure strong interactions between photons. 
Recent developments in the field of 
quantum optics have paved the path in achieving strong non-linearity in various exciting platforms such as the 
optical cavities and superconducting circuits~\cite{Lukinreview_2014_nature_photonics,Hartmanreview, Wallraff2004}. 
 Several path breaking achievements have been made with 
cavity and circuit QED in recent years using the two level artificial atoms(also known as qubits). 
In the many-body context, an array of such artificial atoms coupled by photons have shown to exhibit novel scenarios in the framework of the celebrated 
Jaynes-Cummings-Hubbard(JCH) model~\cite{Hartmann1-2006Nat.Phys.2-849,Hollenberg-L-C-L-2006Nat.Phys.2-856,Makin2008,Mering2009,Schmidt2009,Koch2009,Hohenadler2012,Kenji2013}. The quantum phase transition between the superfluid(SF) and the Mott insulator(MI) 
of polaritons (the quasi-particles composed of atomic excitations and cavity photons) is an important revelation of the competing 
photon-atom interactions inside the cavity 
and the photon hopping between different cavities~\cite{Hollenberg-L-C-L-2006Nat.Phys.2-856}. 
Following this, many interesting quantum phenomena have been analyzed in the framework of the JCH model~\cite{from_non_rectangular_paper}. 
The phenomenal progress in understanding the many-body aspects of strongly correlated photons 
and the demand to fulfill the requirements necessary for quantum technologies have attracted enormous 
attention towards the study of cavity and circuit QED ~\cite{Angelakis1}. Although, primarily the atom-photon interactions in 
such systems are of two-body repulsive and attractive in nature~\cite{Hartmann1-2006Nat.Phys.2-849}, recent progress in manipulating three- and higher level systems 
have provided opportunities to explore novel scenarios in quantum simulations with multi-level systems~\cite{PhysRevLett.123.070505,Guo2019,PhysRevA.83.023814}. 

Although the systems of 
atoms in optical cavities are well established to understand the physics of light-matter interaction, the rapid developments in fabricating superconducting circuits 
have evolved as one of the most suited test bed for quantum simulations in recent years. The versatility of these systems arises from the 
flexibility to control the anharmonicity generated by the Josephson junctions which indirectly controls the interaction between the polaritons~\cite{Schmidt2013}.
Motivated by all the recent developments we analyze the circuit QED of superconducting processor and propose a method to create photon pair propagation. 

In this paper we consider an array of three-level superconducting artificial atoms coupled through microwave resonators as shown in Fig.~\ref{fig:scheme}. In this setup 
the microwave stripline resonator acts as the source of cavity photons and 
the superconducting quantum circuits(SQCs) play the role of three-level artificial atoms. For this purpose we consider the $\Xi$ 
type system with unequal energy spacings as depicted in Fig.~\ref{fig:scheme}. 
The energy difference between the levels $|0\ra$ and $|1\ra$ is denoted as $\varepsilon_{01}$ and between $|1\ra$ and $|2\ra$ as $\varepsilon_{12}$. 
While $\omega$ represents the cavity resonance frequency, $\Delta=\omega-\varepsilon_{12}$, stands for the detuning associated to the 2nd excited level. The many-body physics of this 
system of coupled cavity array(CCA) can be analyzed in the context of the modified JCH model given as; 
\begin{eqnarray}
 \mathcal{H}_{JCH}&=&\sum\limits_i[-\Delta\hat{\sigma}^\dagger_{2i}\hat{\sigma}_{2i}+\beta_{12}(\hat{\sigma}^\dagger_{2i}\hat{a}_i+H.c.)\nonumber\\
 &+&\beta_{01}(\hat{\sigma}^\dagger_{1i}\hat{a}_i+H.c.)]-\kappa\sum\limits_{\langle i,j \rangle}(\hat{a}^\dagger_{i}\hat{a}_{j} + H.c.)
 \label{eq:ham}
\end{eqnarray}
Here, ${a}^\dagger_i({a}_i)$ is the photonic 
creation(annihilation) operator, ${\sigma}^{\dagger}_{1i}({\sigma}_{2i})$ is the atomic raising(lowering) operator which takes  
the atom from $|0\rangle_{i}$ to $|1\rangle_{i}$($|2\rangle_{i}$ to $|1\rangle_{i}$) levels,  
${n}_i={n}^p_i+{\sigma}^\dagger_{1i}{\sigma}_{1i}+{\sigma}^\dagger_{2i}{\sigma}_{2i}$ is the total polariton number at the $i^{th}$ cavity and 
${n}^p_i={a}^\dagger_i {a}_i$ denotes the number operator of the photonic excitations. 
$\beta_{01}(\beta_{12})$ represents the atom-photon 
coupling strength between level $|0\rangle$ and $|1\rangle$($|1\rangle$ and $|2\rangle$). The nearest neighbor inter-cavity photon tunneling amplitude is denoted by $\kappa$. 
For our analysis, we define the polariton density as $\rho=N/L$, where $N=\sum_i n_i$ and L is the total number of polaritons and total number of sites in the system respectively. 
Here we set the cavity resonance frequency $\omega = \varepsilon_{01}$ and assume $\beta_{01}=\hbar=1$ to fix the energy scales.

\begin{figure}[b]
\includegraphics[width=1.\linewidth]{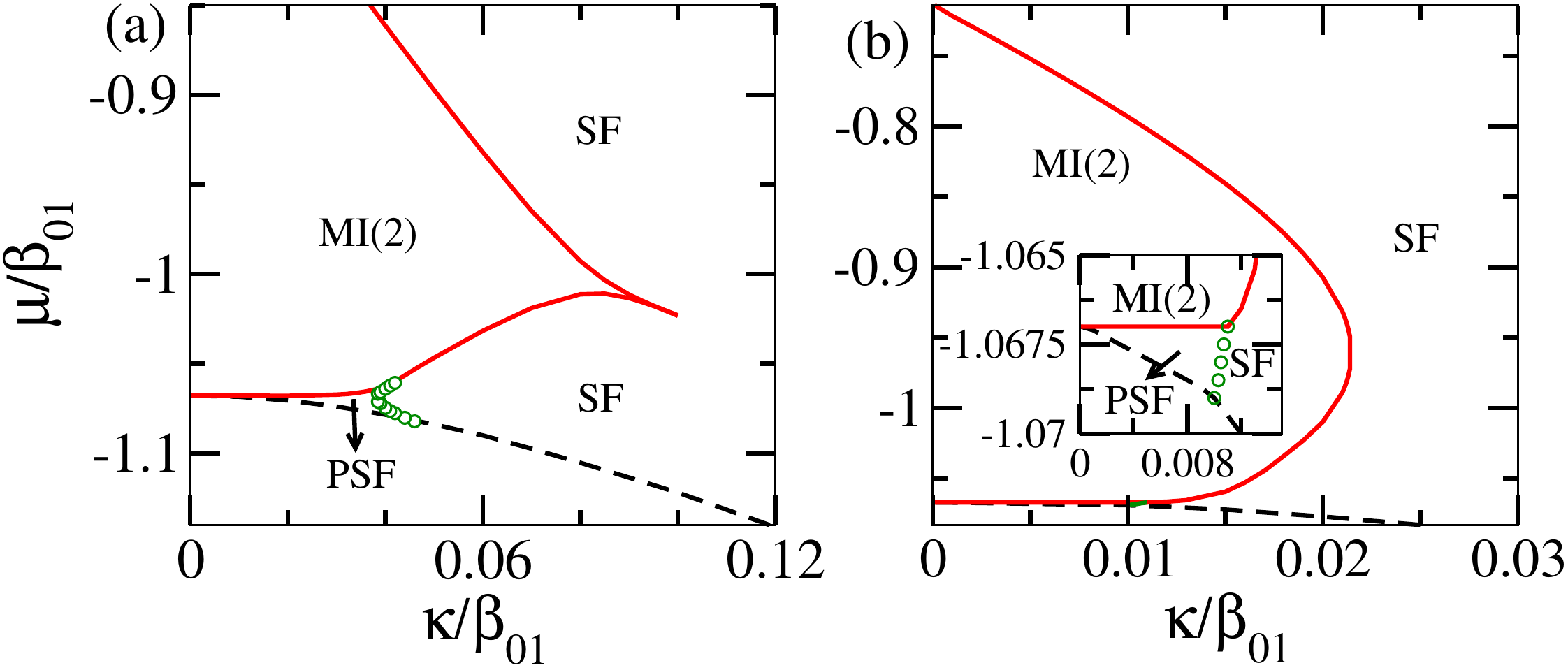}
\caption{(Color online)Phase diagram of the JCH model using (a) the DMRG method in 1$d$ and (b) the CMFT  method in 2$d$ for the detuning $\Delta/\beta_{01} =0.4$. 
In both the figures the red solid curve demarcates the boundary of the MI(2) phase, the green circles show the PSF-SF phase boundary and
 the black dashed curve is the vacuum state or the MI(0) phase. 
 For the DMRG method in Fig.~\ref{fig:superlattice_with_period1}(a) all boundaries are calculated by extrapolating the chemical potential to thermodynamic limit using maximum system size of $L=80$ cavities.
 In the inset of Fig.~\ref{fig:superlattice_with_period1}(b) we show the enlarged PSF region.}
\label{fig:superlattice_with_period1}
\end{figure}

As mentioned before, the two-level JCH model exhibits SF-MI phase transition as a function of the ratio $\kappa/\beta_{01}$. There exist the MI phases at 
integer polariton densities when $\kappa/\beta_{01}$ ratio is small. In the limit $\kappa \gg \beta_{01}$ the MI phases melt and a phase transition to the SF phase occurs due to the delocalization of photons. 
On the other hand a recent mean-field study on three-level atomic system with equally spaced levels in optical cavity arrays predicts the complete 
suppression of the MI(1) lobe after a critical $\beta_{12}/\beta_{01}=\sqrt{2}$~\cite{martin_prasad}. In this limit, the MI(2) lobe is shown to overlap with the vacuum state and this signature
is speculated to be of a PSF phase of polaritons in analogy with the attractive Bose-Hubbard model. 
The key requirement to achieve this phenomenon is that the two transitions, $\ket{0}$ $\rightarrow$ $\ket{1}$ and $\ket{1}$ $\rightarrow$ $\ket{2}$  
should be near resonantly driven by the same photon (including the polarization)  which demands equally spaced three-level $\Xi$ system. 
However, we would like to stress that this condition is not satisfied by the natural atoms in optical cavities.  
Note that for the $\Lambda$ and $V$-systems the frequency of two transitions can be same but requires different polarizations of the photons. 

Interestingly, this condition can be easily satisfied in SQC which plays the role of an artificial atom provided the higher energy levels except the first three are removed or truncated. 
The removal of the higher energy levels can be implemented by using strong anharmonicity to the system~\cite{PhysRevA.76.042319,martinis2002rabi} as depicted in Fig.~\ref{fig:scheme}(b).  
Therefore, in our studies we consider a more realistic system of three-level artificial atoms by considering an SQC with unequal spacings which will 
circumvent the practical issues associated with equal spacing $\Xi$ system. Note that the anhormonicity naturally introduces the detuning for $\ket{1}$ $\rightarrow$ $\ket{2}$ transition. 
To understand the effects of the strong correlations we analyze 
the ground state properties of the model given in Eq.~\ref{eq:ham} for 
one and two dimensional arrays of SQCs using the density matrix renormalization group(DMRG)~\cite{White1992,SchollwockRev,Schollwock2011}
method and the self-consistent cluster mean-field theory(CMFT) approach~\cite{McIntosh,Manpreet2} respectively. 

\begin{figure}[b]
\includegraphics[width=1.\linewidth]{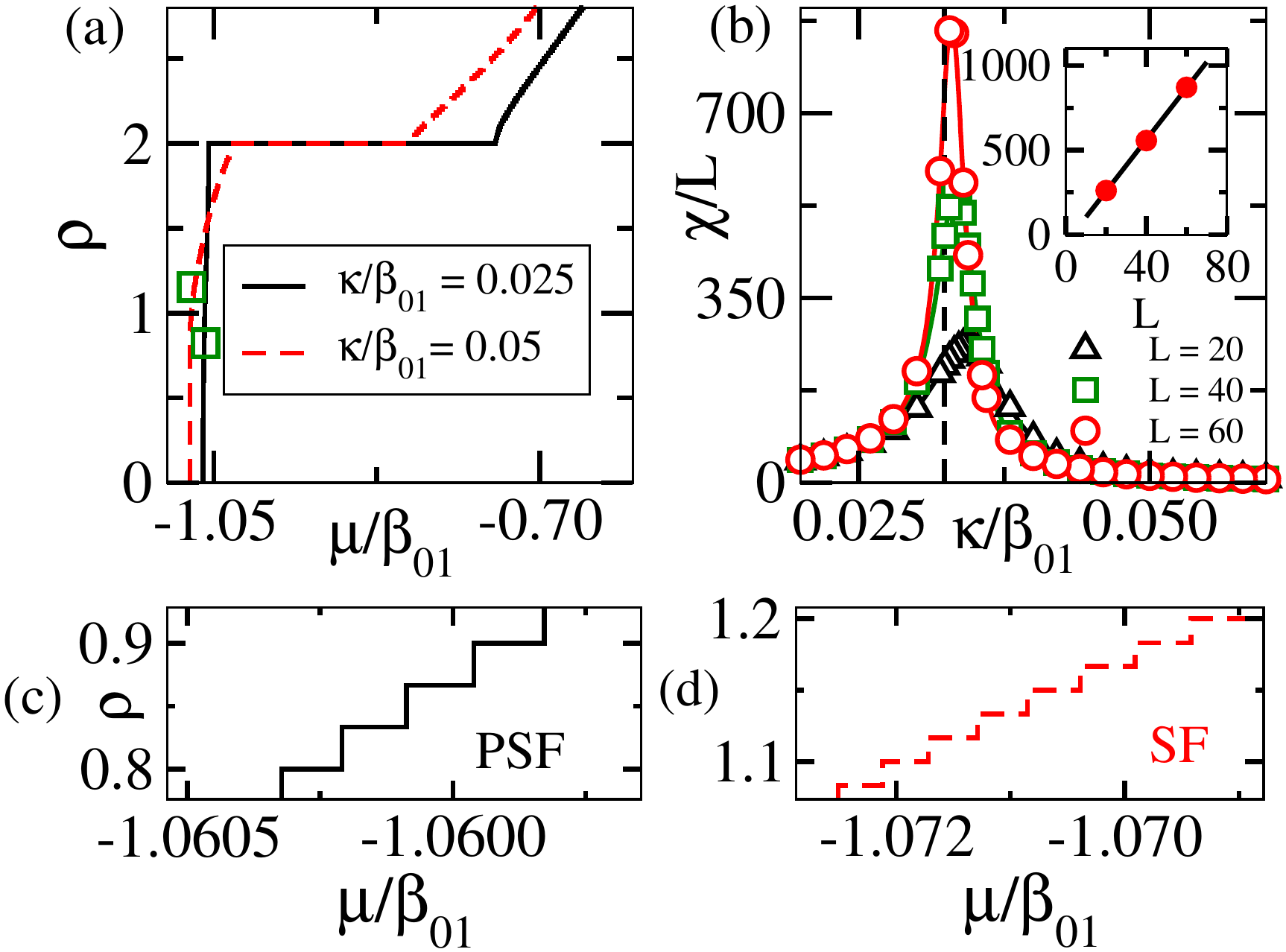}
\caption{(Color online)(a)DMRG data shows the $\rho$ vs $\mu/\beta_{01}$ plot for parameters $\kappa/\beta_{01} = 0.025$ and $\kappa/\beta_{01} = 0.05$ when $\Delta/\beta_{01} = 0.4$ 
indicating the SF and PSF regions for $L=60$ sites. The regions marked by the green boxes are enlarged in Figs.~\ref{fig:rho_mu_fid}(c) and (d) which shows the signatures 
of the PSF and the SF phases respectively.
 (b) $\chi_{FS}(\tilde \kappa)$ vs. $\kappa/\beta_{01}$ plots for different system sizes of $L=20,40$ and $60$ to see the phase transition point.
 (Inset) Shows that the peak heights diverge with system size indicating the phase transition. The dashed vertical line corresponds to 
 the critical point of transition determined by extrapolating the peak position to thermodynamic limit.}
\label{fig:rho_mu_fid}
\end{figure}

{\em Results in 1$d$.-}
In this part we discuss about the results in one dimensional circuit QED array with experimentally 
realistic three-level $\Xi$ system by considering $\beta_{12}/\beta_{01}=\sqrt{2}$~\cite{PhysRevA.76.042319} and finite detuning of $\Delta/\beta_{01} =0.4$. We analyze the effect of 
photon tunneling which couples the SQCs through the capacitive couplings and obtain the ground state properties of the model shown in Eq.~\ref{eq:ham}.  
It is to be noted that there is no particular reason behind this choice 
of $\Delta/\beta_{01}=0.4$. In order to get a clear numerical picture we keep the value of $\Delta/\beta_{01}$ close to the experimentally accessible regime~\cite{PhysRevA.76.042319}. 
By utilizing the DMRG method we compute the ground state phase diagram in the plane of $\kappa/\beta_{01}$ and $\mu/\beta_{01}$ 
as shown in Fig.~\ref{fig:superlattice_with_period1}(a). Note that the DMRG simulations are done in the canonical ensemble with fixed polariton number and hence the Hamiltonian of Eq.~\ref{eq:ham} is 
explicitly independent of $\mu$. It can be clearly seen from the phase diagram that the MI(2) lobe(red solid curve) appears immediately after the vacuum state(black dashed line) 
by completely suppressing the MI(1) lobe which usually appears in the phase diagram of the JCH model of two-level systems~\cite{Hollenberg-L-C-L-2006Nat.Phys.2-856}. 
Moreover, in this case there is no overlap of the vacuum and the MI(2) lobe as opposed to the MFT results shown in Ref.~\cite{martin_prasad} in the absence of any detuning. Interestingly 
there exists a PSF phase of polaritons in the gapless region bounded by the green circles for small values of $\kappa/\beta_{01}$. Before going to the details of this PSF phase
we first discuss about the phase diagram in the following. 

First of all, we trace out the phase transition from the gapped MI(2) phase to the SF phase of polaritons by looking at the energy gaps in the system. 
The signature of the gapped MI(2) phase is seen as the plateaus in the $\rho$ vs $\mu/\beta_{01}$ plot at $\rho=2$ as shown in Fig.~\ref{fig:rho_mu_fid}(a) which is a signature of the gap in the system. 
The phase boundaries are obtained by computing the extrapolated values of the end points of the plateaus which are the chemical potentials of the systems defined as 
$\mu^+=E_{N+1}-E_N$ and $\mu^-=E_N-E_{N-1}$ in the thermodynamic limit for different values of $\kappa/\beta_{01}$. Here, $E_N$ is the ground state energy with $N$ polaritons. 
Now we systematically analyze the signatures of the pair formation in the system. 
The immediate information can be obtained by analyzing the 
dependence of $\rho$ with respect to $\mu/\beta_{01}$ for different values of $\kappa/\beta_{01}$. In Fig.~\ref{fig:rho_mu_fid}(a) we plot $\rho$ vs $\mu/\beta_{01}$ corresponding to
two different values of 
$\kappa/\beta_{01}=0.025$(black solid) and $0.05$(red dashed) of the phase diagram 
in Fig.\ref{fig:superlattice_with_period1}(a). Note that when $\kappa/\beta_{01}=0.05$, the value of $\rho$ increases in steps of one particle, indicating the SF phase. 
However, for  $\kappa/\beta_{01}=0.025$,
the value of $\rho$ increases in steps corresponding to the change in polariton number $\Delta n=2$ up to the MI(2) plateau from the bottom. This can be clearly seen from the zoomed in
regions plotted in Figs.~\ref{fig:rho_mu_fid}(c) and (d) corresponding to the green boxes shown in Fig.~\ref{fig:rho_mu_fid}(a). This indicates the 
quasi particle excitations in terms of polariton pairs which is a typical signature of the pair formation~\cite{manpreetpsf,mishra-greschner_santos,Mondal2019}. This phenomenon 
happens in the gapless region between the vacuum and the MI(2) phase in the regime of small $\kappa/\beta_{01}$ and therefore can be called as a PSF phase of polaritons. 
As a result, there exists a phase transition from the SF phase to the PSF phase as a function of $\kappa/\beta_{01}$ which is indicated 
by the green circles in  Fig.~\ref{fig:superlattice_with_period1}(a). We compute the PSF-SF phase boundary from the $\rho$ vs $\mu/\beta_{01}$ plot and complement it by 
looking at the divergence of the fidelity susceptibility~\cite{gu2010,manpreetpsf} across the phase transition defined as:
\begin{align}
\chi_{FS}(\tilde \kappa) = \lim_{\tilde \kappa-\tilde \kappa' \to 0} \frac{-2 \ln |\langle \Psi_0(\tilde \kappa) |
\Psi_0(\tilde \kappa') \rangle| }{(\tilde \kappa-\tilde \kappa')^2} \,,
\end{align}
at $\rho=1.5$. Here $\tilde{\kappa}=\kappa/\beta_{01}$, $|\Psi_0\rangle$ is the ground-state wave function and $\tilde \kappa'$ is a small change in the rescaled hopping amplitude. 
From Fig. \ref{fig:rho_mu_fid}(b), we observe a diverging stable maximum with increasing system sizes which shows the PSF-SF phase transition point at $\kappa/\beta_{01}= 0.03237$.


Although, the $\rho$ vs. $\mu/\beta_{01}$ behavior 
allows us to identify the PSF phase of polaritons, it does not provide any insight about the underlying mechanism behind this.

\begin{figure}[t]
\includegraphics[width = 1.\linewidth]{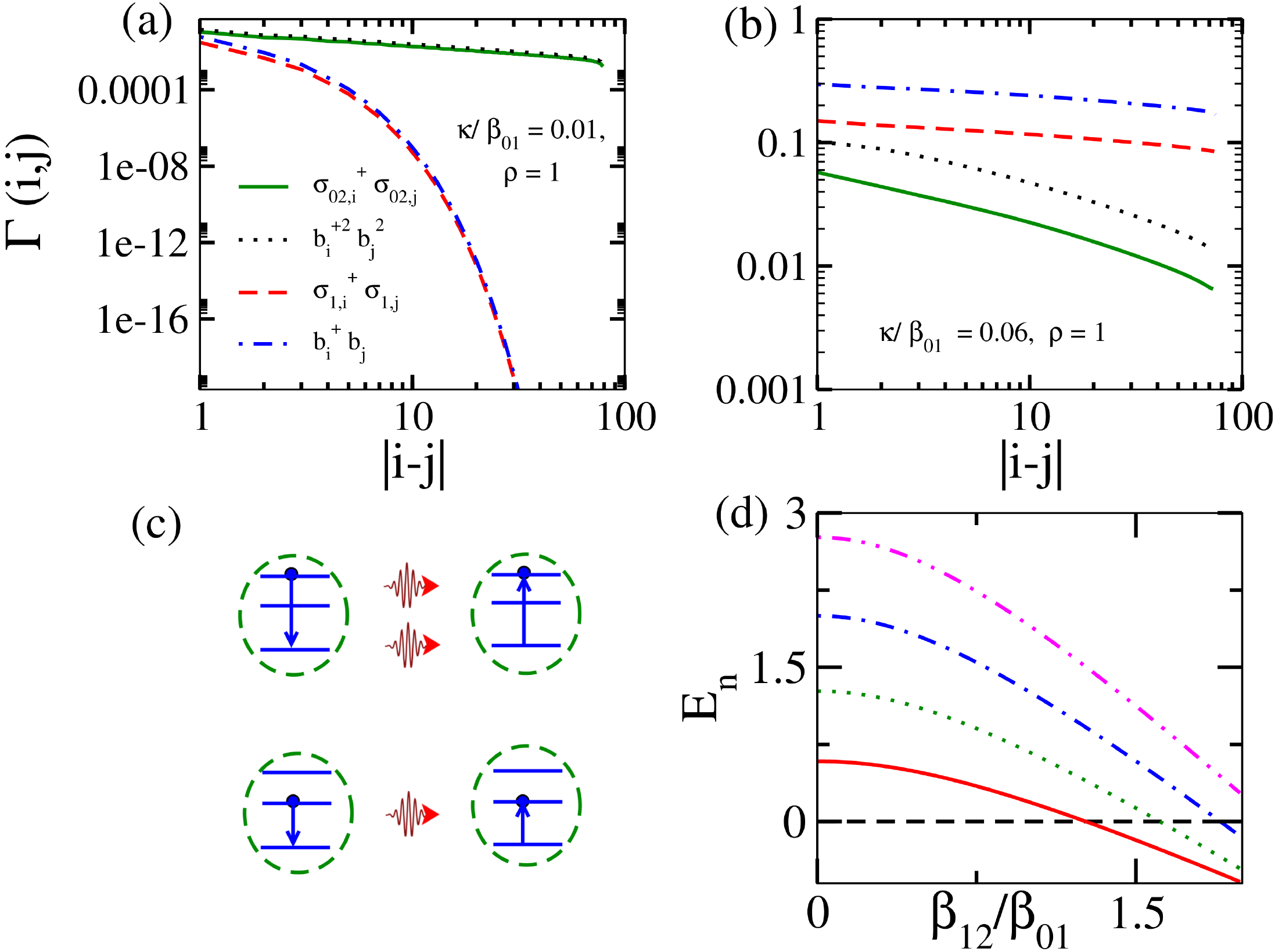}
 \caption{(Color online)Figure shows the pair and single polariton correlation functions $\Gamma(i,j)$ with distance $|i-j|$ 
 for $\rho=1$ with (a) $\kappa/\beta_{01}=0.01$ and (b) $\kappa/\beta_{01}=0.06$. (c) Shows the two photon and single photon tunneling processes. 
 (d)The polariton energies $E_n$(bottom to top for $n=1\to5$ at the origin) for $\mu/\beta_{01}=-1$ with respect to  $\beta_{12}/\beta_{01}$ for $\Delta/\beta_{01}=0.40$.}
 \label{fig:psf_cart}
\end{figure}

{\em The PSF phase.-}
We devote this part of the paper to provide a detailed analysis of the physics of photon pair propagation and the PSF phase of polaritons. To understand the 
pairing phenomena we rely on the behavior of various single and pair correlation functions.  
In Fig.~\ref{fig:psf_cart} we plot all the correlation functions with respect to the distance $|i-j|$ for a system of length $L=80$ and $\kappa/\beta_{01}=0.01$. Interestingly, 
it can be seen Fig.~\ref{fig:psf_cart}(a) that the 
correlation functions associated with the photon pairs which is defined as 
$\Gamma_{photon-pair}(i,j)=\langle b^{\dagger2}_i b_j^2 \rangle$(black dots)  exhibits algebraic decay, where 
as the single photon correlation i.e.  $\Gamma_{photon}(i,j)=
\langle b_i^\dagger b_j \rangle$(blue dot  dashed)  decays exponentially. 
This is a clear indication of the existence of the long-range coherence of photon pairs in the system and the single particle motion is completely suppressed in the 
thermodynamic limit. At the same time the atom-pair correlation defined as $\Gamma_{atom-pair}(i,j)=\langle \sigma_{02,i}^{\dagger}\sigma_{02,j} \rangle$(green solid) also remains finite 
whereas the single atom correlation $\Gamma_{atom}(i,j)=\langle \sigma_{01,i}^\dagger\sigma_{01,j} \rangle$(red dashed) vanishes exponentially across the array. This implies that a pair of photon gets spontaneously 
emitted from a cavity and gets absorbed by the nearest neighbor cavity and excite the atom sitting there. This 
process continues resulting in the superfluid of photon pairs. 
Here  $\sigma_{01,i}(\sigma_{02,i})$ are the annihilation operators associated with the atomic excitations from the ground state to first and second level respectively. On the other hand for large values of $\kappa/\beta_{01}$ we have verified that the 
single particle correlation functions dominate over the pair ones justifying the SF phase as depicted in Fig.~\ref{fig:psf_cart}(b). 
The physical process which may arise from this single and pair photon propagation is depicted in Figs.~\ref{fig:psf_cart}(c). 
Interestingly, we also find that in the limit of two photon propagation
there exists finite correlation corresponding to the single photon and an atomic excitation that is $\Gamma_{atom-photon}(i,j)$. Hence in the present case we can have three different scenarios such as
(a)$|n_p=2,n_a=0 \rangle$, (b)$|n_p=1,n_a=1 \rangle$ and (c)$|n_p=0,n_a=2 \rangle$ which can facilitate the photon pair propagation between the SQCs for small $\kappa/\beta_{01}$ values.

The physics behind such photonic pair creation or photon pair propagation can be understood by analyzing the energies associated to the system as done in Ref.~\cite{martin_prasad}. 
We show in Fig.~\ref{fig:psf_cart}(d) in the presence of $\Delta$ the cavity excitation energy corresponding to two photon becomes negative whereas the energy corresponding to 
other higher polaritonic excitation remains positive well before $\beta_{12}/\beta_{01}=\sqrt{2}$. This promotes the formation of two polaritons in the SQCs and indirectly the photon pair
propagation. Therefore, the photon pair propagation and the associated polaritonic PSF phase in the three-level JCH model is not identical to the atomic PSF phase in the BH model
due to the attractive interaction between bosons.


\begin{figure}[b]
\includegraphics[width=1.\linewidth]{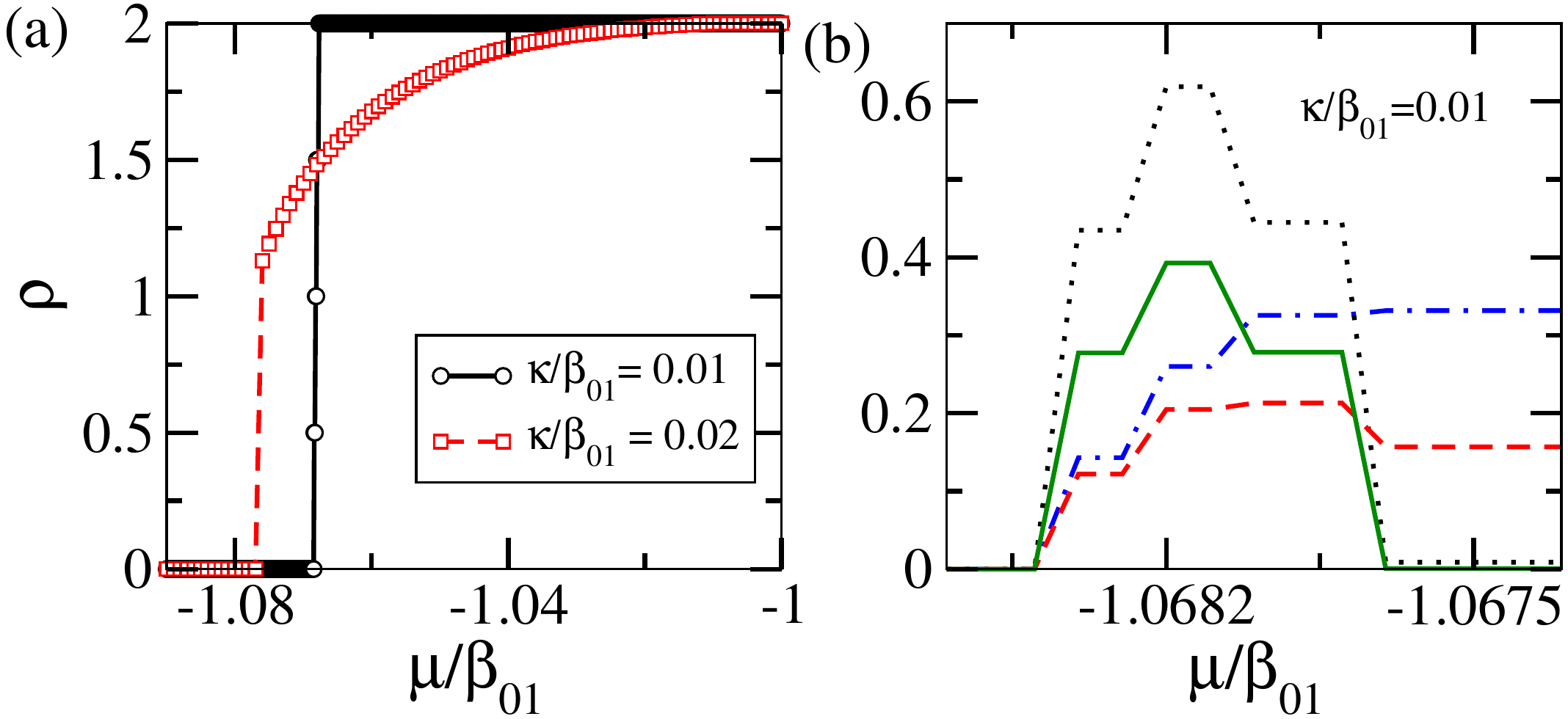}
\caption{(Color online)(a)$\rho-\mu/\beta_{01}$ plot at $\kappa/\beta_{01} = 0.01$ and $0.02$ of Fig.~\ref{fig:superlattice_with_period1}(b).
 (b)Different correlation functions such as $\Gamma_{atom-pair}(i,j)$(Green solid), $\Gamma_{photon}(i,j)$(blue dot-dashed), $\Gamma_{photon-pair}(i,j)$(black dots) 
 and $\Gamma_{atom}(i,j)$(red dashed) are plotted for $\kappa/\beta_{01} = 0.01$ (see text).}
\label{fig:rho_mu_corr_cmf}
\end{figure}

{\em Phase diagram in $2d$.- }After obtaining the signature of the photon pair propagation in the one dimensional circuit QED setup 
we analyze the physics of the JCH model using the CMFT approach by going to two dimension. Note that the CMFT approach works in the grand canonical ensemble 
and hence we explicitly include the term associated to the chemical potential as $\mu \sum_i n_i$ in the JCH model given in Eq.~\ref{eq:ham}.  
In this method the entire system is divided into identical clusters of limited number of sites which can be treated exactly 
and then the coupling between different clusters are 
treated in a mean-field way. The accuracy of this method improves by increasing the number of sites in the cluster. 
With this approximation the original Hamiltonian of Eq.~\ref{eq:ham} can be written as 
\begin{eqnarray}
 \mathcal{H}_{CMF}&=&\mathcal{H}_{C}+\mathcal{H}_{MF}\nonumber\\
 &=&\mathcal{H}_{C}-\kappa \sum\limits_{\langle i,j \rangle}[(a^\dagger_i+a_i)\psi_j-\psi_i^* \psi_j]
 \label{eqn:4}
\end{eqnarray}
where, $\mathcal{H}_C(\mathcal{H}_{MF})$ is the cluster(mean-field) part of the Hamiltonian and $\psi_i={\langle{a}^\dagger_i\rangle}={\langle{a}_i\rangle}$ is 
the SF order parameter. The form of $\mathcal{H}_C$ is same as Eq.~\ref{eq:ham} and is limited to the cluster only. 
The self consistent solution of the CMFT Hamiltonian yields the ground state phase diagram in two dimension as depicted in Fig.~\ref{fig:superlattice_with_period1}(b). 
It can be seen that the phase diagram in 2$d$ 
is qualitatively similar to the one obtained for the 1$d$ case (Fig.~\ref{fig:superlattice_with_period1}(a)). The phase diagram of Fig.~\ref{fig:superlattice_with_period1}(b)
is obtained by looking at the behavior of the density 
$\rho= \frac{1}{L}\sum_i\hat{n}_i=\frac{1}{L}\sum_i(\hat{n}^p_i+\hat{n}^a_i)$ with respect to $\mu$ for different values of $\kappa/\beta_{01}$.  
In Fig. \ref{fig:rho_mu_corr_cmf}(a) we plot the values of $\rho$ vs. $\mu/\beta_{01}$ along the cuts through the CMFT phase diagram of Fig. \ref{fig:superlattice_with_period1}(b) 
at $\kappa/\beta_{01} = 0.01$  and $\kappa/\beta_{01} = 0.02$ which pass through different phases.
The discrete jumps in the $\rho-\mu/\beta_{01}$ plot (black circles) in steps of two particles is an indication of the PSF phase 
as discussed before and the plateaus at $\rho = 2$ corresponds to the MI(2) phase. 
We also plot the correlation functions for a single photon, a pair of photons, single atom and a pair of atoms as shown in Fig. \ref{fig:rho_mu_corr_cmf}(b). As the cluster is of four sites only, 
the correlations are computed between the nearest neighbors and averaging them over the entire cluster.
This clearly shows the dominant pair correlation functions as compared to the single particle ones(see figure caption for details). 

{\em Conclusions.-} In this work we propose a scheme for spontaneous photon pair creation and propagation in an array of coupled SQCs. Considering 
the three-level artifical atoms of $\Xi$ type instead of the usual two-level qubit systems we analyze the corresponding Jaynnes-Cummings Hubbard model 
in one and two dimensional arrays using the DMRG and the CMFT approach to establish the emergent photon pair propagation in the system. 
We show that for the suitable ratio of the coupling strengths between different levels, the single photon tunneling is suppressed and photons tend to move in pairs. 
This two photon propagation leads to the 
formation of polaritonic pair superfluid phase which is located in between the vacuum and the MI(2) phases of the polaritonic phase diagram. This finding is obtained by 
considering a more realistic setup of the SQCs of three-level atom with unequal level spacings which is experimentally more feasible as opposed to the optical cavity-atom setups. 
We would like to note that in this case, there exists no overlap between the vacuum state and the MI(2) phase or the first order type phase transition as predicted earlier 
using the MFT approach~\cite{martin_prasad}. 
This inconsistency can be attributed to the artifact of the simple 
mean-field theory approach using which it is difficult to capture all the relevant physics arising due to the off-site correlations as rightly mentioned in Ref.~\cite{martin_prasad}.

This analysis provides a promising platform to observe the pairing phenomena of bosons in general as compared to its atomic and molecular counterparts. 
Moreover, this finding in the three level system can possibly be made useful for quantum communications~\cite{PhysRevLett.123.070505,Guo2019,PhysRevA.83.023814} 
in the future as bound state of photons is believed to carry more information than the individual photon. 
This work can shed light on the controlled creation and manipulation of boson pairs and can be extended to create 
higher order photonic bound states(trimers etc.) in an array of multi-level artificial atoms.

\begin{acknowledgments}
T.M. acknowledges DST-SERB for the early career grant through Project No.  ECR/2017/001069. K.P. acknowledges funding from SERB of grant no. ECR/2017/000781.
Part of the computational simulations were carried out using the Param-Ishan HPC facility at the Indian Institute of Technology - Guwahati, India. 
\end{acknowledgments}

\bibliography{jch_model_psf_ref}
 
\end{document}